# Estimation of Rectifying Performance for Terahertz Wave in Newly Designed Fe/ZnO/MgO/Fe Magnetic Tunnel Junction

H. Saito, and H. Imamura

*National Institute of Advanced Industrial Science and Technology (AIST) Spintronics Research Center, Central 2, 1-1-1 Umezono, Tsukuba 305-8568, Japan*

We fabricated fully epitaxial Fe/ZnO/MgO/Fe magnetic tunnel junctions (MTJs) with low junction resistance-area products (several $\Omega\mu m^2$) and conducted a theoretical estimation of square-low rectifying performance for a terahertz electromagnetic wave. Effective current responsivity up to 0.09 A/W at 1 THz was obtained under zero-bias voltage condition at room temperature. The result is approximately half the value of the best result obtained for experiments in semiconductor-based diodes, performed under similar conditions. The study strongly suggests that this MTJ system has a great potential for the rectifying element of the terahertz wave.

## 1. Introduction

Studies concerning terahertz (THz) wave, having a frequency ($f$) range from 100 GHz to 10 THz, have recently emerged in expectation with the great potentials for novel applications such as ultra-fast wireless communications[1-3] and non-destructive imaging.[4,5] A THz wave receiver is one of the crucial building blocks of such THz applications[6]. However, developments of high-performance THz devices including the receiver are still challenging issues because using the corresponding frequencies leaves a "gap" between those for conventional electronics and optics system. Low manufacturing cost is also an important requirement from a practical device point of view.

Metal-insulator-metal (MIM) tunnel diodes are attractive rectifying element for the THz/far infrared applications[7-14] in this respect. Although the MIM diode has been generally considered as possessing poor rectifying performance when compared with semiconductor-based diodes, it has the advantage of realizing very low junction resistance-area products ($R_DA$) of below 1 $\Omega\mu m^2$ at a zero-bias voltage. This leads to the shorter RC time constant, and thereby resulting in higher cut-off frequency of the system, compared to the semiconductor-based diodes. Moreover, the MIM diode is suitable for mass-production due to its feasibility of employing conventional manufacturing growth methods, such as a sputtering, even on a flexible substrate[15] at low temperatures.

Very recently, we found a strong diodic behavior in the current-voltage ($I-V$) characteristics in epitaxial MI(I)M device of Fe(001)/ZnO(001)/MgO(001)/Fe(001) magnetic tunnel junction (MTJs) at room temperature (RT)[16], in which tunneling magnetoresistance (TMR) is caused by so called spin-polazied coherent tunneling[17,18]. We theoretically estimated the effective current responsivity ($\beta_{eff}$) of the MTJ system, which correnponds to the conversion capability from a high-frequency to a DC current, in regards to the ZnO and MgO barrier thicknesses ($t_{ZnO}$ and $t_{MgO}$, respectively) as well as the junction area ($A$)[19]. The calculation predicted that $\beta_{eff}$ can reach up to 0.12 A/W at 1 THz with the conditions, $t_{ZnO} \sim t_{MgO}$ with low $R_DA$ (a few $\Omega\mu m^2$) and small $A$ (~ 0.01 $\mu m^2$). The obtained $\beta_{eff}$ is comparable to that of the best result obtained in semiconductor-based diodes in experiments for InP/(In,Ga)As system, performed under similar conditions (~ 0.2 A/W at 1 THz [20,21]). However, most of the physical parameters of the tunnel barriers used in our previous calclation[19], such as electron affinity and relative dielectric constant, were extracted from the experimetal $I-V$ data of the MTJ having high $R_DA$ (an order of k$\Omega\mu m^2$) with large $A$ (36 $\mu m^2$)[16]. These values are actually three orders higher than desirable values for effective rectification of the THz wave. The next important step, therefore, is to fabricate and characterize the electrical properties of the MTJ with comparatively low $R_DA$ and small $A$, and evaluate the feasibility in such cases.

In this study, we fabricated the epitaxial Fe(001)/ZnO(001)/MgO(001)/Fe(001) MTJs with $R_DA$ of several $\Omega\mu m^2$ and $A \sim 0.01$ $\mu m^2$. Based on a simple antenna-coupled diode model[7], we theoretically estimated $β_{eff}$ of the MTJ prepared in this study, showing that it is possible to achieve the $β_{eff}$ as high as 0.09 A/W at 1 THz under a zero-bias voltage at RT.

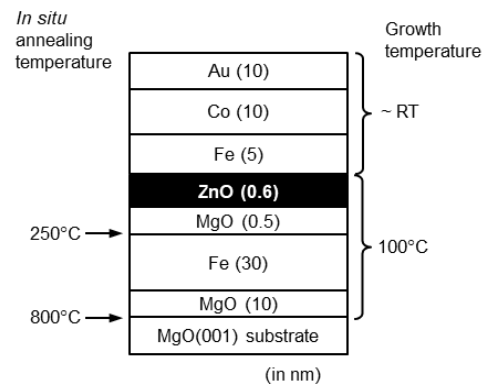

**Fig. 1** Structure of the magnetic tunnel junction (MTJ) stack designed for this study. Growth and *in situ* annealing temperatures are also shown.

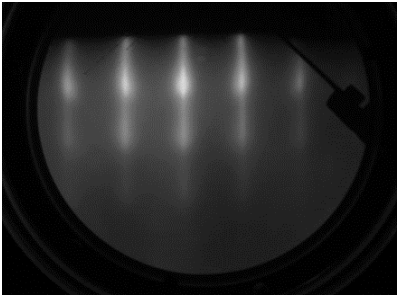

**Fig. 2** Reflection high-energy electron diffraction image of the rock-salt type ZnO tunnel barrier ([100] azimuth of MgO substrate).

## 2. Experimental procedures

A MTJ film as shown in Fig. 1 was grown by molecular beam epitaxy in the identical growth chamber as used for the growth of epitaxial MTJs in our previous studies[16,20]. The MTJ film consisted of Au (5 nm) cap / Co (10 nm) pinned layer / Fe (5 nm) top electrode / ZnO (0.6 nm) upper tunnel barrier / MgO (0.5 nm) lower tunnel barrier / Fe (30 nm) bottom electrode / MgO (5 nm) buffer layer on a MgO(001) substrate, which is so called pseudo-spin valve structure. The $t_{ZnO}$ and $t_{MgO}$ were determined, in accordance with our theoretical calculation[19], so as to realize the maximum $β_{eff}$ in this MTJ system. The ZnO upper tunnel barrier was confirmed to have identical reflection high-energy electron diffraction patterns (Fig. 2) to that of the MgO(001) lower tunnel barrier, indicating a single-crystalline with a metastable rock-salt type crystal structure. Detailed growth procedures and structural properties were described in our previous study.[18]

The film was patterned into tunnel junctions having elliptical shape using electron beam lithography technique. The lengths of the major and minor axes of the junctions were respectively 150 and 50 nm, corresponding to $A$ = 0.0059 μm². Magnetoresistance (MR) and $I$–$V$ characteristics of the junctions were measured at RT using conventional DC four and two probe methods, respectively. The MR ratio was defined as $(R_{AP}–R_P)/R_P$, where $R_P$ and $R_{AP}$ are the junction resistances between the two ferromagnetic (FM) electrodes with parallel (P) and antiparallel (AP) alignments, respectively. The magnetic fields were applied parallel to the major axis of the junction corresponding to the easy axis of the magnetization direction of the FM electrodes.

## 3. Results and discussions

### 3.1 Electrical transport measurements

Figure 3 shows a typical MR curve of the MTJs at RT. The observed MR ratio (43%) is close to the experimental value of epitaxial Fe/ZnO/MgO/Fe MTJs (51%) with thicker tunnel barrier layers ($t_{ZnO}$ = 1.2 nm and $t_{MgO}$ = 1.0 nm), grown at the same growth conditions in our previous study.[18] The MR curve always showed a low resistance state (i.e. the P state) at a zero-magnetic field, indicating that there is no significant magnetic coupling between the top and bottom FM electrodes. In Fig. 4, the first derivatives of the current (d$I$/d$V$) is displayed with regard to the bias voltage in the P state. The data were obtained from numerical differentiation of the $I$–$V$ characteristics. The d$I$/d$V$–$V$ curve indicated a notable asymmetry with respect to the polarity of the bias voltage, suggesting that the MTJ has a capability of square-low rectification at a zero-bias voltage.

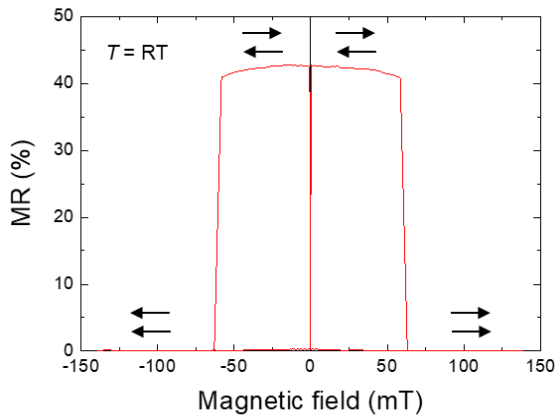

**Fig. 3** Typical magnetoresistance (MR) curve of the Fe/ZnO/MgO/Fe MTJ at room temperature (RT) with a bias voltage of 10 mV. Arrows indicates the magnetization alignments in the top and bottom electrodes

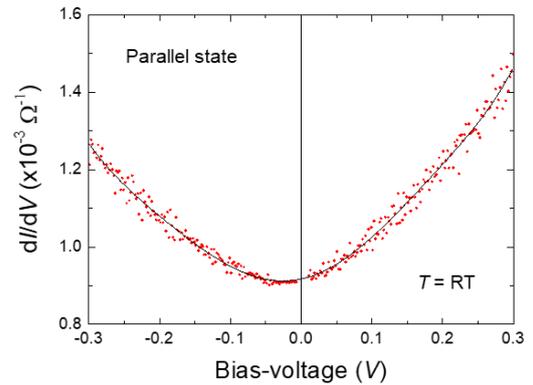

**Fig. 4** (Red circles) Bias-$V$ dependence of d$I$/d$V$ of the Fe/ZnO/MgO/Fe MTJ at RT in the parallel (P) state of the magnetizations of the top and bottom electrodes. (Solid line) Polynomial fitting result to the d$I$/d$V$ data.

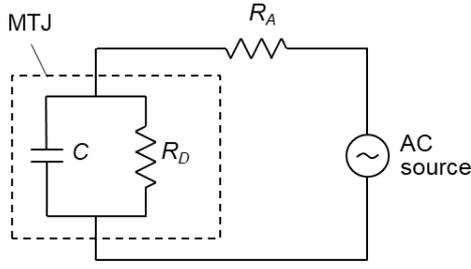

**Fig. 5** Equivalent circuit of the antenna-coupled diode model[7] used in this study.

To calculate $\beta_{\text{eff}}$, we obtained two important parameters of $R_DA$ [$\equiv$ $(dI/dV)^{-1}A$] and $\beta_0$ [$\equiv (1/2)(d^2I/dV^2)/(dI/dV)$] from polynomial fittings to the $dI/dV$–$V$ curve at a zero magnetic field (i.e. in the P state). Here, the parameter $\beta_0$ is equal to $\beta_{\text{eff}}$ only when the resistance of the MTJ ($R_D$) completely matches with the internal real impedance ($R_A$), and also when the $f$ of the incident electromagnetic wave is far below the cut-off frequency of the system. Therefore, $\beta_0$ does not necessarily represent the rectifying performance of the diode although it is in general referred to as the performance index. The obtained $R_DA$ (6.3 $\Omega\mu m^2$) and $\beta_0$ (0.20 A/W) were in good agreements with the theoretically estimated values in the identical MTJ ($R_DA$ = 8.9 $\Omega\mu m^2$ and $\beta_0$ = 0.23 A/W[19]), indicating that the MTJ prepared in this study justifies the design concept, and also fabricated precisely in line with the concept for the THz rectification.

### 3.2 Estimation of rectifying performance

The $\beta_{\text{eff}}$ under a high-frequency condition is given by the following equation based on the antenna-coupled diode model (Fig. 5) [7],

$$\beta_{\text{eff}} = I_{dc}/P_{in} = 4\beta_0 x[1+2x+(1+q^2)x^2]^{-1}, \quad (1)$$

where $I_{dc}$ is the rectified DC current, $P_{in}$ is the power of incident electromagnetic wave. The variable $x$ is the reduced junction resistance (= $R_D/R_A$), and $q$ is the reduced frequency (= $2\pi fCR_A$). Note that $x$ represents an impedance matching between the junction and the internal impedance whereas $q$ is related to the RC time constant of the system. For $R_D$ and $\beta_0$, we employed the experimental values in Fig. 4 ($R_D$ =1.1 k$\Omega$ and $\beta_0$ = 0.20 A/W), so that the calculation was conducted in the P state. Inverse of junction capacitance ($C^{-1}$) for the present MTJ is given by $(\varepsilon_0\varepsilon_{ZnO}A/t_{ZnO})^{-1} + (\varepsilon_0\varepsilon_{MgO}A/t_{MgO})^{-1}$, where $\varepsilon_0$ is the dielectric constant, and $\varepsilon_{ZnO}$ and $\varepsilon_{MgO}$ are the relative dielectric constants of rock-salt ZnO (12[16]) and MgO (8.8[23]), respectively. Consequently, $\beta_{\text{eff}}$ can be determined when $f$ and $R_A$ are given. Other possible factors of propagation or energy losses in high frequencies were not taken into account so that the calculated $\beta_{\text{eff}}$ are the upper bounds in the MTJ.

The calculated $\beta_{\text{eff}}$ are summarized in Figs. 6 and 7. As shown in Fig. 6, each $\beta_{\text{eff}}$–$R_A$ curve took a maximum at a certain $R_A$ which corresponds to the matching condition between $R_D$ and $R_A$. For $f$ = 100 GHz, a maximum $\beta_{\text{eff}}$ (0.19 A/W) was approximately equal to $\beta_0$, indicating that the cut-off frequency is higher than 100 GHz. We confirmed no remarkable change in the $\beta_{\text{eff}}$–$R_A$ relation below $f$ = 100 GHz. When further increasing $f$, $\beta_{\text{eff}}$ starts to decrease, especially in a high $R_A$ regime. This is because the contribution of the RC time constant (the parameter $q$) to $\beta_{\text{eff}}$ becomes significant in the high $R_A$ regime. It would be possible to improve the cut-off frequency and thereby increase $\beta_{\text{eff}}$ by reducing the $R_DA$ down to a few $\Omega\mu m^2$ [19]. Nevertheless, $\beta_{\text{eff}}$ with the impedance matching condition still has a sizable value of 0.09 A/W at 1 THz which is approximately half the value of the best result obtained in experiments for semiconductor-based diodes (~ 0.2 A/W at 1 THz), performed under similar conditions[20,21]. Note that, as mentioned above, the $\beta_{\text{eff}}$ calculated in the present MTJ are the upper bounds. Therefore, from practical viewpoint, it is crucial to exhibit maximum rectifying performance by minimizing the propagation or energy losses in high frequencies as much as possible.

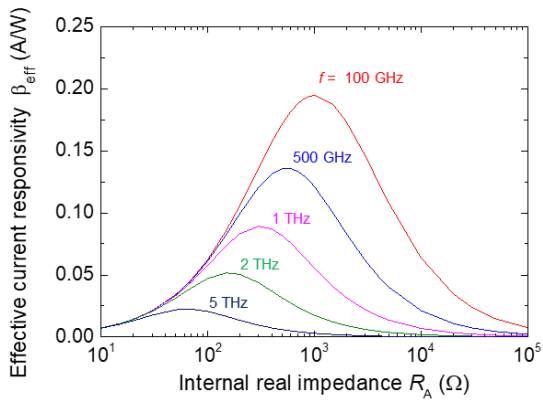

**Fig. 6** Calculated effective current responsivity ($\beta_{\text{eff}}$) as a function of internal real impedance ($R_A$) for the Fe/ZnO/MgO/Fe MTJ prepared in this study. The values are plotted for various frequencies ($f$) of the input electromagnetic wave. The calculations were conducted at a zero-bias voltage in the P state.

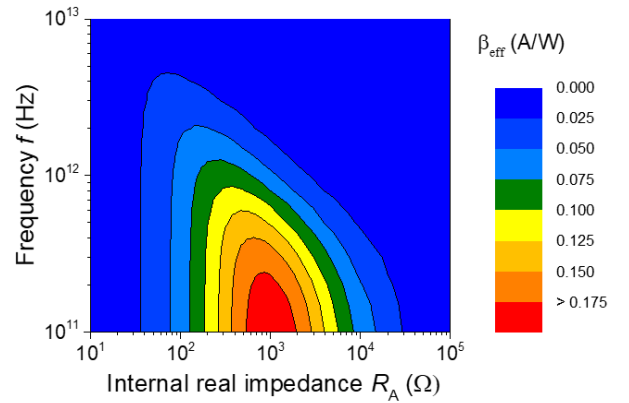

**Fig. 7** Calculated $\beta_{\text{eff}}$ of the Fe/ZnO/MgO/Fe MTJ as functions of $f$ and $R_A$.

It should be noted that the rectified DC current and voltage in MTJ depend on the relative angle between the magnetizations of the two FM electrodes according to the changes in $R_D$ and $\beta_0$. The phenomenon as such however, is impossible to accomplish in semiconductor-based diode and conventional MIM tunnel diode with a non-magnetic electrode(s). In other words, MTJ-based MIM diode provides novel functionality regarding manipulation of the rectified current/voltage by magnetic field in addition to the traditional rectifier. This unique function of MTJ may give rise to new device applications in high-frequency technology.

## 4. Conclusion

We fabricated fully epitaxial Fe/ZnO/MgO/Fe MTJs and theoretically estimated $\beta_{eff}$ in a terahertz regime as a square-low detector. We obtained $\beta_{eff}$ up to 0.09 A/W at 1 THz under a zero-bias voltage condition at RT, which is almost half the value of the best result obtained in experiments for semiconductor-based diodes, performed under similar conditions. The results strongly suggest that this MTJ system has a great potential for a rectifying element of a terahertz wave.

**Acknowledgements** We thank Dr. Akio Fukushima (AIST) for device fabrications and fruitful discussions. We also thank Dr. Hitoshi Kubota (AIST), and Dr. Masahiro Horibe (AIST) for valuable discussions.